# Phosphorescence by trapping defects in boric acid induced by thermal processing

Luigi Stagi[1*], Luca Malfatti[1], Alessia Zollo[2], Stefano Livraghi[2], Davide Carboni[1], Daniele Chiriu[3], Riccardo Corpino[3], Pier Carlo Ricci[3], Antonio Cappai[3], Carlo Maria Carbonaro[3], Stefano Enzo[4], Abbas Khaleel[5], Abdulmuizz Adamson[5], Christel Gervais[6], Andrea Falqui[7], and Plinio Innocenzi[1,5*]

[1] Laboratory of Materials Science and Nanotechnology, CR-INSTM, Department of Biomedical Sciences, University of Sassari, Viale San Pietro 43/B, 07100 Sassari, Italy

[2] Department of Chemistry and NIS, University of Turin, Via P. Giuria 7, 10125 Turin, Italy

[3] Department of Physics, University of Cagliari. Sp 8, km 0.700, 09042 Monserrato, CA, Italy

[4] Department of Chemical, Physics, Mathematics and Natural Sciences, University of Sassari. Via Vienna 2. 07100 Sassari. Italy

[5] College of Science, Department of Chemistry. United Arab Emirates University. Al Ain. United Arab Emirates

[6] Sorbonne Université, CNRS, UMR 7574, Laboratoire de Chimie de la Matière Condensée de Paris, LCMCP, F-75005 Paris, France

[7] Department of Physics "Aldo Pontremoli", University of Milan, Via Celoria 16, 20133 Milan, Italy

Corresponding author email: lstagi@uniss.it ; plinio@uniss.it

Keywords: boric acid; phosphorescence; borates; electron traps; defects

## ABSTRACT

The phosphorescence of boric acid at room temperature is a puzzling phenomenon subject to controversial interpretations although the role of structural defects has not yet been considered. Heat treatments of boric acid cause its transformation into the metaboric phase and amorphous boron oxide. The structural changes after thermal processing can create defects that become centers of luminescence and recombination channels in the visible range. In the present work, we have thermally processed commercial boric acid at different temperatures. Samples treated between 200 and 400°C exhibit remarkable phosphorescence in the visible range. At around 480 and 528 nm, two distinct phosphorescent emissions occur, associated with trapped charge carriers recombinations identified by thermoluminescence and electron paramagnetic resonance spectroscopy. Our structural and optical studies suggest that the activation of boric acid phosphorescence after heat treatment is correlated with the presence of defects. The afterglow results from a trapping and detrapping process, which delays the recombination at the active optical centers. Time-dependent density functional study of defective BOH molecules and clusters shows the emergence of near UV and blue optical transitions in absorption. These defects trigger the photoluminescence in thermally processed boric acid samples.

## 1. Introduction

Hydrogen borate or boric acid (BA, $H_3BO_3$) is a compound of significant interest in the glass industry. It can be synthesized by reacting borax ($Na_2H_{20}B_4O_{17}$) with hydrochloric acid and typically exists in the form of colorless crystals or white powders.[1]

Upon heating at temperatures higher than 100 °C, boric acid undergoes multiple dehydration steps and ultimately forms boron oxide ($B_2O_3$).[2,3] This process is accompanied by structural transformations and melting until a vitreous oxide is obtained. At room temperature (RT), this vitreous oxide may incorporate the hydroxide phase as a result of either incomplete dehydration or exposure to atmospheric aqueous vapor.[4]

Despite their wide range of potential applications, the role of boric acid and its oxide in materials obtained through heat treatment in the air or solvo/hydro-thermal processing has yet to be fully understood. Recent studies have explored the use of boric acid in combination with organic compounds to create fluorescent and phosphorescent materials.[5–10] Boric acid, as a co-precursor in the synthesis, promotes the condensation of organic compounds into graphitic/graphene structure with enhanced fluorescence yield and significant redshift. Furthermore, when harnessed in a boric acid matrix or in polymers with boric acid, the resulting phosphors show a strong quenching resistant fluorescence and, in some cases, a solid state phosphorescence.[5] For instance, mixing organic precursors and boric acid can lead to the production of phosphorescent carbon dots with blue and green emissions and glowing lifetimes of 0.1-1 s after high-temperature treatment.[7,10] The main interpretation of the appearance of phosphorescence is the creation of a strong covalent boron-carbon bond along with the effective role played by the boron oxide glassy matrix in altering the vibrational and rotational characteristics of the organic moieties.[7] It is interesting to note that in all the mentioned cases the phosphorescence occurs in solid-state, where the structural (crystalline or amorphous) properties of the BA matrix are preserved, without purification to remove BA species. The emission primarily falls in the blue (450 – 500 nm) and green (520 – 550 nm) optical regions and the excitation channels between 300 and 400 nm. The mechanism of phosphorescent recombinations is usually attributed to spin-forbidden relaxation from triplet states in a not well-defined electronic structure of the phosphors. For example, room-temperature phosphorescence (RTP) has been recently observed in commercial boric acid and attributed to recombinations from triplet states. Zheng and coworkers[11] have found that commercial boric acid powders exhibit a long afterglow in the blue region, with a maximum at 450 nm. At RT, the recombination lifetime is in the order of a tenth of a second and reaches about 1 s at 77 K. This effect has tentatively been attributed to a weak conjugation among n electrons of O atoms in packed BA molecular units, although the simulated vertical transitions fail to indicate a possible absorption in the near ultraviolet (NUV) that is responsible for the blue phosphorescence. This explanation has been critically revised by Wu et al.,[12] who have compared the luminescence of commercial $H_3BO_3$ with that of trimethyl borate $(B(OMe)_3)$-derived boric acid, which is impurity-free. Under UV irradiation, they have not observed phosphorescence or optical recombination in pure boric acid although they have observed luminescence in commercial boric acid. The source of the emission in commercial samples has been associated with the existence of impurities, as the boric acid produced from $B(OMe)_3$ was free from impurities.[12] These findings suggested that, in addition to its established features, boric acid should have novel and unexplored luminescent properties with potential implications for solid-state platforms. To fully exploit these phosphorescent characteristics and effectively design phosphorescent materials, it is of paramount importance a comprehensive study of the structural properties of boron oxide and hydroxide and their inherent properties as a function of thermal processing. Investigating the properties of boric acid as a function of processing temperature up to its oxidation stage is a crucial strategy to understand the role of crystallographic phases and defects that may occur during a synthesis involving boric acid. This would enable us to discriminate the possible role of boron and its coordination with oxygen from any other interaction with a heteroatom (C, S, N etc.) and allow the attribution of any phosphorescent properties to specific electronic states of the B-O structure.

In this work, we investigate the origin of phosphorescence in commercial boric acid treated at various temperatures. Specifically, boric acid exhibits blue and green RTP under UV light excitation when treated near the melting point. However, this RTP quenches after high temperature treatments, which mainly result in the conversion of boric acid into boron oxide ($B_2O_3$). Furthermore, the RTP can be restored by dissolving $B_2O_3$ in water, evaporating the water to reestablish the boric acid structure, and then retreating the material. These findings suggest that the RTP observed in boric acid arises from structural defects.

## 2. Results and Discussion

Commercial boric acid has been heated at various temperatures before being cooled to room temperature for characterization (**Figure 1a**). During a standard calcination process, BA powders have been placed in a ceramic crucible and subjected to heating under air atmosphere for a duration of 2 h at the selected temperature. The resulting material exhibits distinct morphological properties depending on the specific temperature employed, with BA transforming into either a coarser, grainy powder or a more densely packed vitreous solid. Within a limited range of temperatures, the treatment of pristine powders produces blue and green phosphorescent samples whose afterglow can be easily appreciated by the naked eye (**Figure 1b**).

It is well established that the crystal and amorphous structures of boric acid can vary depending on its oxidation state.[13] Thermogravimetric analysis (TGA-DSC) has been carried out over a wide temperature range to assess the relevant temperatures required to induce structural transformations. Our results show that during heating, BA undergoes distinct stages of dehydration and melting. **Figure 1c** shows the combined TGA and DSC analysis of $H_3BO_3$ performed under nitrogen flux from room temperature up to 800 °C with a heating rate of 10 °C $\cdot$ min$^{-1}$. Two main endothermic events are observed in the range between 80 and 200 °C. The first DSC event has been recorded at 130 °C and corresponds to a dehydration process of boric acid promoting the formation of the *orthorhombic metaboric acid* phase:

$$H_3BO_3 \rightarrow HBO_2 + H_2O \quad (80 - 130°C) \tag{1}$$

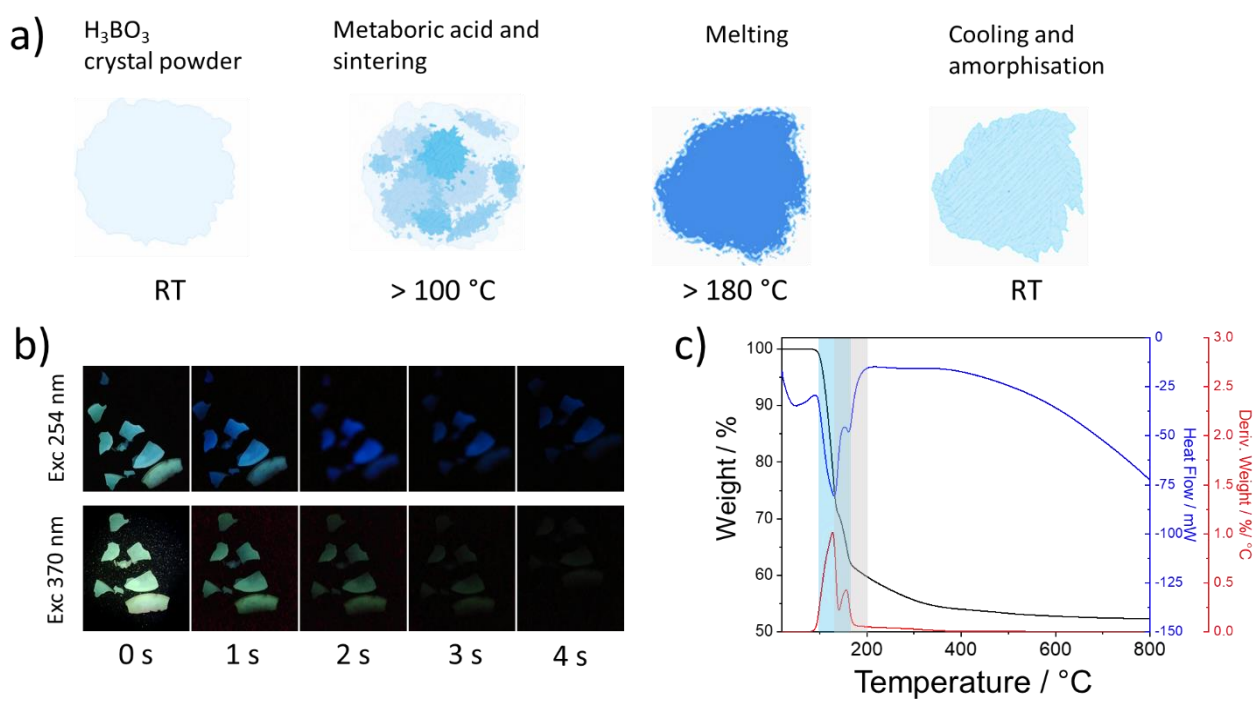

**Figure 1**. a) Artistic picture of boric acid thermal treatment performed in this work. b) Images of phosphorescent samples (BA250) acquired with a camera at 1 s intervals after two different excitation wavelengths; 0 s corresponds to the instant of switching off the source. c) DSC-TGA analysis of as purchased $H_3BO_3$. Grey and light blue areas highlight the temperature interval between the two dehydration processes.

In this range, the sample appears sintered and forms a rough bulk solid. This first process is followed by a second dehydration, where metaboric acid oxidizes into $B_2O_3$ (*boron trioxide*):

$$2HBO_2 \rightarrow B_2O_3 + H_2O \quad (144 - 210 \text{ °C}) \tag{2}$$

Tetraboric acid ($H_2B_4O_7$) is assumed to occur in the second dehydration process between 180 and 200 °C, although there is no clear evidence of an associated stable phase after cooling.[14]

According to TGA curves, almost 30 % of weight loss is associated with the first dehydration event, while an abrupt change in slope is observed during the second one. The oxidation progresses with the temperature at a slower pace. The lack of further thermal events suggests completion of BA melting and low-rate desorption of volatile species.

In **Figure 2a** we report the XRD patterns of the as-purchased $H_3BO_3$ sample, and after thermal treatment in air at increasing temperatures. The samples are therefore named BA, BA100, BA130, BA200, BA300, BA400 and BA800. The $H_3BO_3$ diffractogram measured at RT shows a perfect match with that of crystalline boric acid (JCPDS 30-0199, sassolite). BA is characterized by $B(OH)_3$ molecules interacting by hydrogen bonds and arranged in a planar configuration (boxes in **Figure 2b**). The corresponding XRD pattern displays the two main reflections at 14.6° and 28.2°, which originate from (010) and (002) lattice planes. Detailed patterns are separately reported in **Figure S1**.

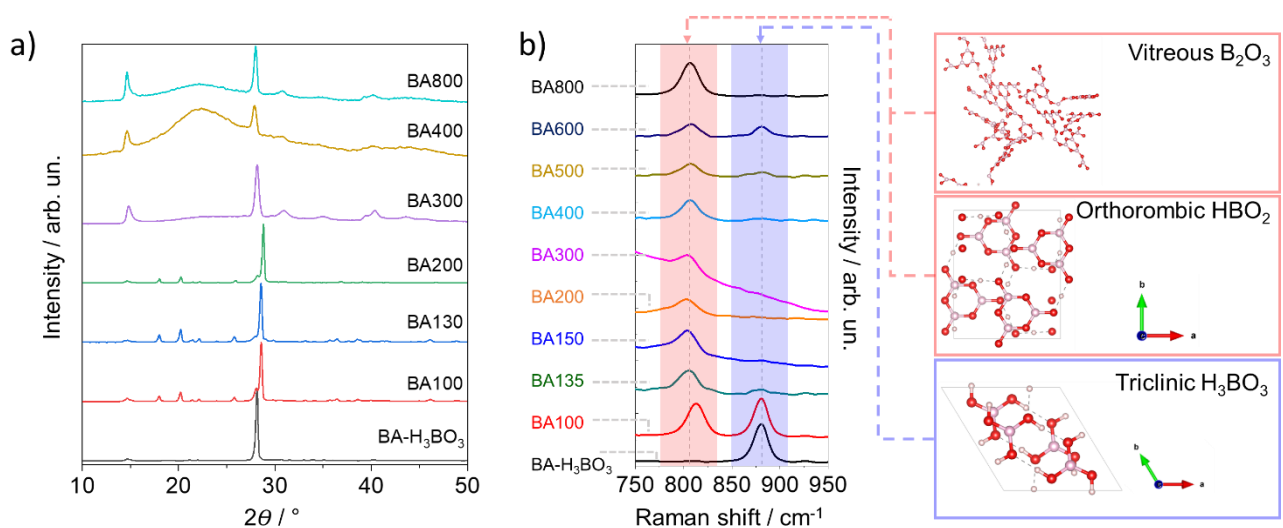

**Figure 2**. a) X-ray diffraction patterns of pristine $H_3BO_3$ powder and thermally treated BA at 100 (BA100), 130 (BA130), 200 (BA200), 300 (BA300), 400 (BA400) and 800 °C (BA800). b) Raman spectra of thermally treated BA powders in the range between 750 and 950 $cm^{-1}$. Characteristic OH bending in $H_3BO_3$ (881 $cm^{-1}$) and breathing mode in $B_3O_3$ boroxol ring (807 $cm^{-1}$) are highlighted. Pictures of the structural phase identified by combining XRD and Raman spectroscopy data are reported in blue and red boxes.

Upon heating the boric acid sample to 100°C, it completely converts to orthorhombic metaboric acid (JCPDS 77-0425), $HBO_2$, with a residual $H_3BO_3$ (28%). However, at 130°C, the relative amount of $HBO_2$ increases to 77% and reaches 87% at 200°C. After higher temperature treatments, the samples cool down in a mixture of $H_3BO_3$ and amorphous boron oxide. The identification of the phase composition and crystallite size has been determined using Rietveld analysis (**Figure S1**).

Boroxol rings are interconnected by $BO_3$ tetragonal units, as illustrated in **Figure 2b**. Boron oxide tends to convert into boric acid when exposed to water-absorbing conditions. This phenomenon is responsible for boron oxide instability.[4] This research has revealed a further important factor to take into account, namely a partial recrystallization of boric acid after melting.

The molecular arrangement of treated samples has been investigated by vibrational spectroscopies. Raman spectroscopy measurements in the range between 750 and 950 cm$^{-1}$ have been reported in **Figure 2b**. Full spectra of pristine $H_3BO_3$ and BA800 are included in **Figure S2**. Eight vibrational modes are identifiable in boric acid under visible light excitation ($\lambda_{exc}$ = 532 nm). The most intense Raman line is observed at 881 cm$^{-1}$ and is attributed to OH bending in tetragonal BOH.[15] Some $H_3BO_3$ modes are still observable in BA800 in accordance with XRD data, although the most intense line is the breathing mode of boroxol rings at 807 cm$^{-1}$.[16] BA100 is characterized by the presence of tetragonal BOH and boroxol rings (metaboric acid structure). After increasing the temperature, the boroxol ring mode becomes predominant. Raman spectra have been collected in the temperature treatment range of 135 to 400 °C by exciting at 1064 nm to minimize fluorescence, which is still observable in BA300. Accordingly, in this interval, the samples exhibit a boroxol ring structure that can be attributed both to the metaboric acid and $B_2O_3$ vitreous structures. Although with different occurrence, $B(OH)_3$ line at 881 cm$^{-1}$ is always present demonstrating the partially recrystallisation as well as the conversion of oxide structure into hydroxide.[17]

**Figure 3a** shows the FTIR absorption spectrum of as purchased $H_3BO_3$ in the 4000-400 cm$^{-1}$ range. The most intense vibrational mode is located at 1464 cm$^{-1}$ and corresponds to B-O stretching $\nu(BO)$. We also observe the B-O bending $\beta(BOH)$ at 1192 cm$^{-1}$ and B-O stretching $\nu(BOH)$ at 888 cm$^{-1}$.[15] An intense and broad absorption is recorded at 800 cm$^{-1}$ and is attributed to B-OH out-of-plane bending $\beta_{out}(BOH)$. At 640 and 540 cm$^{-1}$ we measure a further B-OH out-of-plane bending $\beta_{out}(BOH)$ and an O-B-O antisymmetric bending $\beta(OBO)$.[15] The FTIR absorption spectra as a function of temperature is shown in **Figure 3b**. The spectra have also been recorded *in situ* (in KBr pellets) as a function of the temperature from 25 up to 200°C. Heat treatment of boric acid promotes a marked modulation of infrared absorption, with a broadening of the bands accompanied by a shift towards lower wavenumbers. The resulting spectrum assumes absorption characteristics detectable in numerous $B_2O_3$-based glasses.[18–21] We can identify at least three broad bands in the region between 2000 and 600 cm$^{-1}$ as a result of the thermal treatment. In borate glasses, the absorption in the range 1500-1220 cm$^{-1}$ is attributable to $BO_3$ stretching.[21] In particular, the two components with maxima at 1398 and 1291 cm$^{-1}$ arise from B-O stretching vibration of borate rings and boroxol rings.[22] Small contributions from the B-O$^-$ stretching may also be present around 1520 cm$^{-1}$. The least intense band at 1084 cm$^{-1}$ lies in the BO$_4$ stretching region of tri-, tetra- and pentaborate groups. The third region is dominated by strong absorption at 720 cm$^{-1}$ due to the B-O-B bending of oxygen bridges between two boron atoms, one of which is in a trigonal configuration between two boroxol ring units.[22]

The formation of an interconnected boron oxide structure proceeds with the temperature and the first change is observed around 120-140°C, at the beginning of the second dehydration stage when, in accordance with DSC-TGA findings, boric acid condenses into metaboric acid. Dehydration can be monitored in the 4000-2750 cm$^{-1}$ range, which displays how the B-OH stretching mode of H$_3$BO$_3$ (3200 cm$^{-1}$) varies as the temperature rises (**Figure 3c**). At around 160°C, we observe a second band peaking around 3600 cm$^{-1}$, partially overlapped with the 3200 cm$^{-1}$ mode whose intensity diminishes with the increase in temperature. This band is tentatively assigned to B-OH stretching in isolated and non-hydrogen bonded OH species.[23] Boric acid at 25°C is a well-known hydrogen bonded solid, and the broad nature of the 3200 cm$^{-1}$ band is due to the stretching of B-OH species in a hydrogen bonded environment. With the proceed of dehydration and the subsequent condensation that gives boron oxide, the length of hydrogen-bonded B-OH chains decreases and more likely remains isolated B-OH, responsible for the signal at 3600 cm$^{-1}$. This finding can be better visualized using a 3D map of the FTIR spectra, wavenumber (x-axis), temperature (y-axis) and intensity (false colour scale) (**Figure 3d**). Stage I, from 25 up to around 130°C, corresponds to the first dehydration phase; stage II is an intermediate stage where the first and second dehydrations simultaneously occur. Finally, stage III corresponds to the second dehydration with condensation and the formation of an oxide network with the production of B-OH dangling hydrogen species.

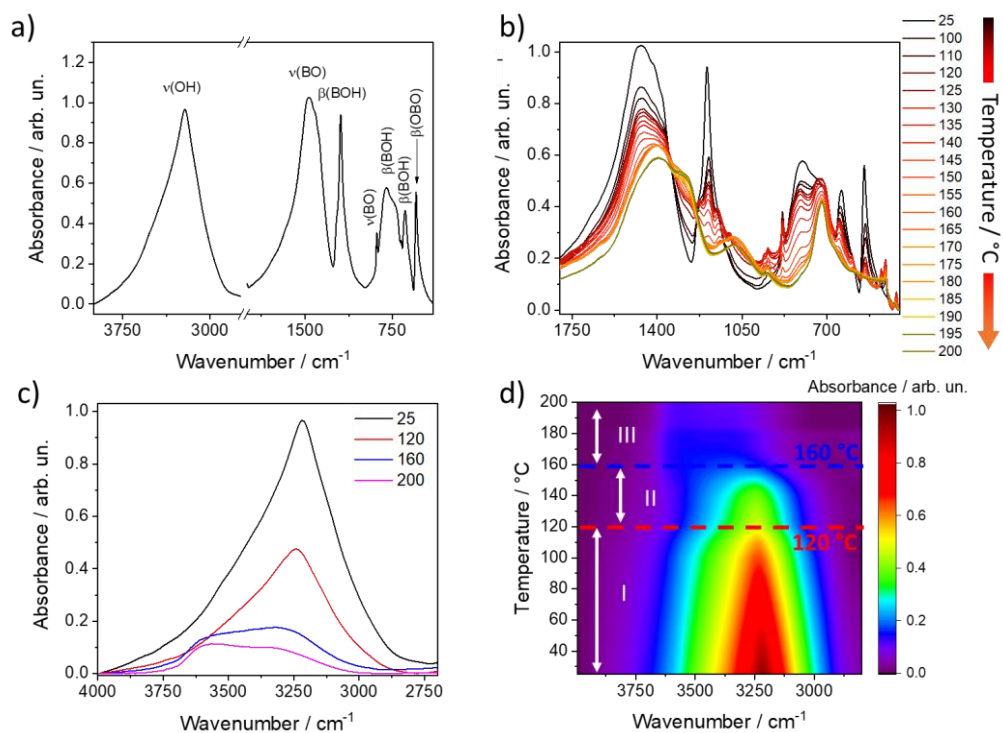

**Figure 3**. a) FTIR absorption spectrum of as purchased $H_3BO_3$ in the 4000-450 cm$^{-1}$ range. b) FTIR absorption spectra in the 1775-950 cm$^{-1}$ range, recorded in situ at temperatures ranging from 25 up to 200°C. c) FTIR absorption spectra in the 4000-2750 cm$^{-1}$ range, recorded in situ at temperatures ranging from 25 up to 200°C. d) 3D map of the FTIR spectra (2800-4000 cm$^{-1}$), wavenumber (x-axis), temperature (y-axis) and intensity (false colour scale). The FTIR map indicates three different stages in the process, identified as I, II, III.

It is worth mentioning that *in situ* infrared analysis monitors the characteristics of the sample at a given temperature in KBr matrix. As mentioned before, the actual structure of the sample could be slightly different when cooled down at RT. This can be observed in conventional FTIR spectra at RT (**Figure S3**). $^{11}$B MAS NMR spectrum of boric acid (**Figure S4a**) shows a well resolved second order quadrupolar line shape that can be fitted with parameters ($\delta_{iso}$ = 19.2 ppm; $C_Q$ = 2.6 MHz; $\eta_Q$ = 0.2) characteristic of $BO_3$ trigonal environment in agreement with previously reported results.[24,25] After heat-treatment at 250°C, spectrum of BA250 sample is significantly changed: $BO_3$ peak remains observable, albeit with slight modifications, indicating a deviation from perfect trigonal symmetry, possibly due to factors like partial deprotonation. Furthermore, an additional narrower signal is observed at 0.5 ppm that can be assigned to the appearance of $BO_4$ tetragonal environment.[26] $^1$H MAS NMR spectra of boric acid and BA250 (**Figure S4b**) show signals centered at 8.5 and 6.7 ppm respectively, indicating modification of the proton environment during heat-treatment and possible protonation of the $BO_4$ sites. After higher heat-treatment, $^{11}$B MAS NMR spectrum of BA800 shows only $BO_3$ signals and complete disappearance of $BO_4$ ones.

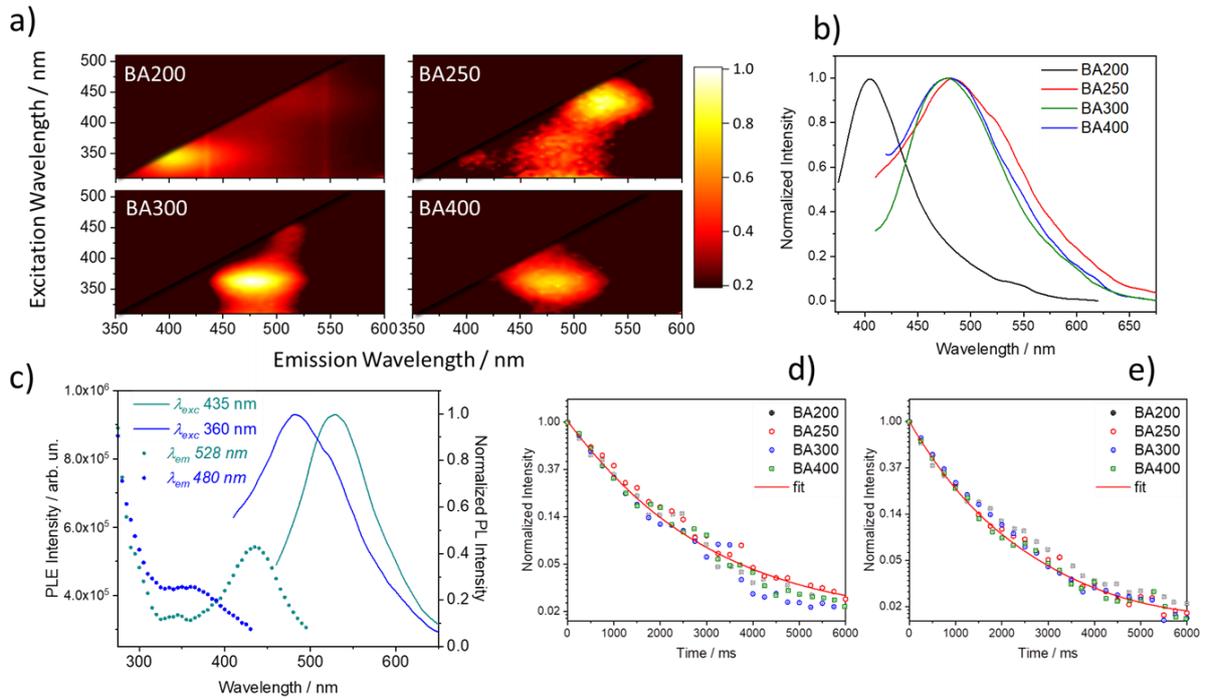

**Figure 4.** a) Excitation-Emission spectra (x-axis, emission wavelength; y-axis, excitation wavelength; z-axis, intensity emission in false colours) of BA200, BA250, BA300 and BA400. b) Normalized emission spectra under excitation at 360 nm. c) Excitation ($\lambda_{em}$ = 435 nm; 528 nm) and emission spectra ($\lambda_{ex}$ = 360 nm; 480 nm) of BA250. Time-resolved spectra of the blue emissions at 410 nm (BA200) and 480 nm (BA250, BA300, BA400) (d) and green emissions at 528 nm (BA200, BA250, BA300, BA400) (e).

In **Figure 4** we report a summary of the optical properties of BA samples treated at different temperatures. In our experimental setup for the acquisition of steady-state photoluminescence, only samples heated to temperatures between 200 and 400 °C emit detectable luminescence. PL-PLE analysis (**Figure 4a**) shows that BA samples are characterized by a blue (~410-480 nm) and a green (~528 nm) component with a slight difference in position and intensity according to the treatment temperature. After temperature treatment at 200 °C, BA emits in the near UV-blue region with a peak at around 410 nm (**Figure 4a**), excitable at 360 nm. A faint emission is also observed at about 520 nm (**Figure 4a**). The temperature treatment at 250 nm enhances the green emission and redshifts the blue one. This effect is accompanied by a growth of the green component over the blue emission peaking at 460 nm. In **Figure 4c**, we report an example of excitation channels in the range between 250 and 500 nm for BA250 (a spectrum extended to <250nm is reported in **Figure S5**). Both components are excitable below 250 nm, where the band-to-band absorption tail is located. At higher wavelengths, the excitation is more selective, with maxima at 360 and 435 nm for the emissions at 480 and 528 nm, respectively. After further temperature increases to 300 and 400 °C, the 460 nm emission becomes the dominant one. In addition, the sample at 300 °C appears to be the most efficient under the same experimental conditions. The emission decays in **Figures 4d** and **4e** were assessed using two-exponential decays, yielding recombination lifetimes of 0.632 ms and 1.656 ms for the blue emission, and 0.420 ms and 1.278 ms for the green emission (**Table S1**).

In accordance with previously reported findings,[27] the sample morphology changes as a result of heat treatment. The boric acid fine powder undergoes a form of sintering process characterised by larger grain sizes, compacting the sample, and causing it to firmly adhere to a substrate (**Figure S6a**). At low processing temperatures (< 200 °C), the sample is still flaky and easily reducible to a powder. At higher temperatures, the glassy appearance becomes dominant, and the sample crumbles with greater difficulty or sticks to the substrate more tightly.

In another experiment, boric acid has been placed on a silicon substrate and heated to 200 °C for 1 h. By means of Raman spectroscopy, we have detected regions with different structural properties. Point analysis exhibits regions of phase inhomogeneity, i.e. areas with a higher contribution of boroxol rings and areas with residual contributions of the $B(OH)_3$ structure. This underlines the difficulty of removing residual OH at higher temperatures (**Figure S6b**).[28]

$H_3BO_3$ synthesized sample (here indicated as BA-$H_3BO_3$*) has been obtained by reacting boron trichloride solution ($BCl_3$) with water and subsequently recrystallized to collect a crystalline powder (**Figure S7**). BA-$H_3BO_3$* shows identical structural and optical properties to the commercially available BA sample. Moreover, it exhibits the distinctive phosphorescence characteristic after treatment at 250 °C, confirming that this effect is inherent to boric acid and is not influenced by the specific synthesis procedure. Both commercially sourced and laboratory-synthesized samples have also been subjected to micro X-ray fluorescence analysis, confirming the absence of specific impurities introduced during synthesis and thermal processing (**Figure S8**).

From studying the heating process of boric acid, we can summarise the general trend of B-O coordination from room temperature to 800 °C. At temperatures above 100 °C, the trigonal planar $B(OH)_3$ coordination of metaboric acid quickly converts into the trigonal $HBO_2$ structure with residual $H_3BO_3$. $HBO_2$ exhibits an O-B-O molecular structure in the form of boroxol rings with possible tetrahedral units coordinated to the boron atom ($HBO_2(II)$, $HBO(I)$).[28,29] At higher temperatures, boroxol rings tend to arrange in disordered structures with bridging boron atoms in trigonal coordination. Based on our structural and optical measurements, the onset of phosphorescence can be correlated with the formation of boroxol rings, but it cannot be univocally assigned to a specific structural phase.

According to their electronic structures, boric acid and high temperature derived structures are in fact not expected to display luminescent recombination in the visible range. The electronic properties of $H_3BO_3$ as well as $HBO_2$ were investigated by DFT calculations.[30] Bulk triclinic boric acid 2A exhibits an indirect band gap of 5.98 eV. As a result of hydrogen bond-type interactions between in-plane units and van der Waals forces between out-of-plane units, the molecular nature dominates over the long-range structure. Considering the molecular orbitals, HOMO and LUMO predominantly involve OH groups.[30] Similarly, metaboric acid-based structures show a bandgap above 6 eV and an indirect to direct gap transition is observed for the α-orthorhombic polymorph. It is also interesting to note the behaviour of the orbitals involved when boroxol structures are formed. In this case, the electronic excitation originates from the p-type orbitals of the oxygen (HOMO) and then involves the π-type character of the $B_3O_3$ ring. This effect is shared by the orthorhombic (α-I) and monocline (β-II) structures, while the cubic (γ-III) structure has no actual participation of boron atoms in the LUCB (Lowest Unoccupied Conduction Band).[31] This evidence shows that the electronic properties of boric acid, the various polymorphs of metaboric acid and even boron oxide in its glassy and crystalline form cannot explain the presence of radiative recombinations in the visible spectrum of any sort, fluorescence, or phosphorescence.

We hypothesize that the origin of thermally induced phosphorescence in BOH-based structures can be ascribed to structural defects. In addition, the sample can be treated at high temperatures to quench the phosphorescence, dissolved in water, and reprocessed until the phosphorescence emission is restored without an obvious difference, which supports the intrinsic origin of emissions.

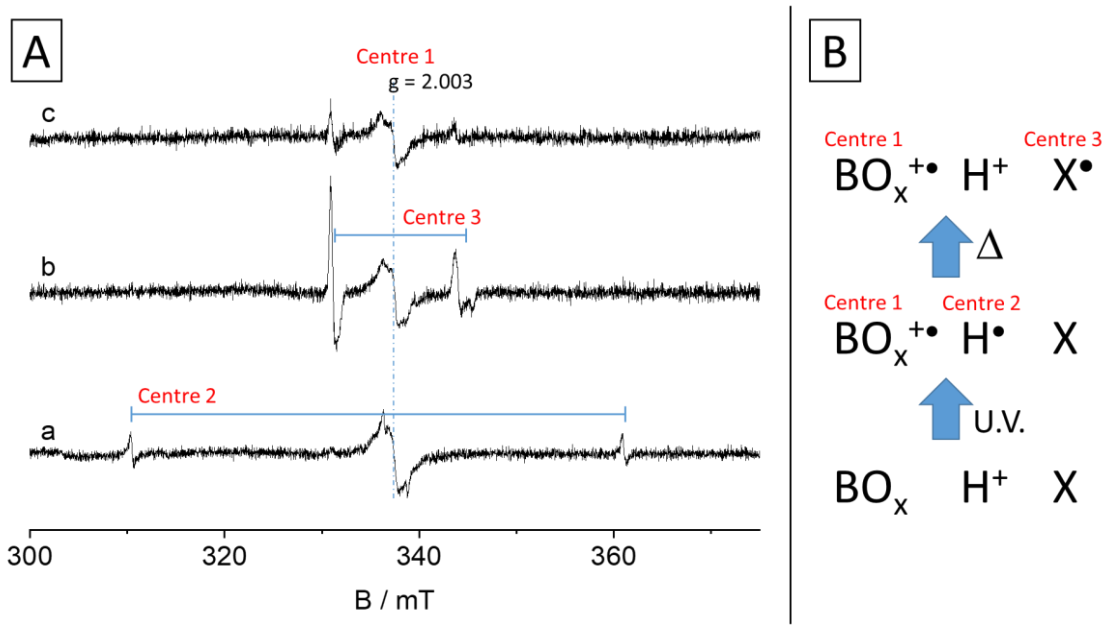

**Figure 5:** Panel A: EPR spectra of BA250 sample a) under UV irradiation b) after exposing the EPR cell to room temperature c) after one day at room temperature. Panel B: sketch of the processes related to the spectra evolution in panel A and involving a Boron containing structural defect ($BO_x$), an isolated proton and a further not assigned defect X.

To confirm the presence of structural defects, EPR characterization has been performed. In **Figure 5**, panel A illustrates the EPR spectra acquired at 77 K while irradiating BA250 with UV light. Upon irradiation, a distinct spectrum, labelled "*a*", emerges, revealing two discernible signals. The first signal corresponds to a paramagnetic species centered at g = 2.003 (designated as Centre 1). The spin-resonance lines appear broad and unresolved, but it is still feasible to identify spectroscopic characteristics indicative of hyperfine coupling with a boron nucleus, approximately ranging from 0.7 to 0.8 mT. (**Figure S9**). Such a signal is compatible with boron centred radical as Boron Oxygen Hole Centres (BOHC) or Boron Electron Centres (BEC).[32] In aluminoborate glasses, BOHC defects can entrap holes on an oxygen bridging trigonal and tetragonal boron atoms. On the other hand, it is widely acknowledged that the defects in BEC defects arise from the entrapment of electrons at an unoccupied site of a nonbridging oxygen within $BO_4$ units.[32] Similar hyperfine coupling are also reported for borate anions impurities in calcium carbonates.[33] A second isotropic signal centred at g = 2.002 split in two components, is also detected (Centre 2). The observed splitting results from the coupling with a nucleus possessing a spin quantum number I = ½. In our context, this can be attributed exclusively to a hydrogen atom. This paramagnetic center can be identified as an isolated proton radical (H•).[34] The hyperfine coupling for this species is indeed 50.5 mT, indicating that the spin density is almost 100 % localized on hydrogen nucleus as calculated from the spin density distribution formula for an *s*-type orbital ($c_s^2 = A_{iso}/A^*_{iso}$ where $c_s^2$ = spin density distribution, $A_{iso}$ = isotropic splitting observed and $A^*_{iso}$ = isotropic splitting in the free atom, 50.8 mT for the proton).[35] The process leading to the formation of this paramagnetic centre is described by the first step of the sketch in **Figure 5B**. After turning off the lamp the two species are stable, until exposure to room temperature (spectrum *b*), after which the hydrogen radical signal disappears and simultaneously a new paramagnetic centre (Centre 3) arises. The spin Hamiltonian parameters for this last species, obtained by computer simulation, are $g_\perp$ = 2.003, $g_{//}$ = 1.996 and $A_\perp$ = 12.8 mT, $A_{//}$ = 13.9 mT. The hyperfine tensor **A** indicates that also in this case a hydrogen nucleus is involved in the paramagnetic centres. However, in this case the spin density on the proton accounts for only about 26%. The generation of this last centre can be rationalized as a transfer of the spin density from the hydrogen radical to a neighbouring defect (second step of the scheme in **Figure 5B**). Since no hyperfine coupling with boron is observable in Centre 3, it is reasonable to assign it to an oxygen centred radical. Even after one day at room temperature, the signals persist, although they appear less intense compared to their initial intensity (spectrum *c*). If the same experiment is carried out using visible light only ($\lambda$ > 400 nm) no signals are detected indicating that the observed behaviour is strictly related to the UV light. Finally, it is worth mentioning that pristine $H_3BO_3$ sample shows a completely different behaviour (**Figure S10**). Under irradiation, only Centre 1 can be observed in pristine boric acid. This suggests that only the centres 2 and 3 have a role in the luminescence of the material. While uncertainties remain regarding the identification of the defective $BO_x$ center, the EPR characterization undeniably reveals the existence of structural defects that respond to light irradiation, involving radical oxygens and species associated with atomic hydrogen, with a strong dependence on temperature.

The trigonal boron ($BO_3$) units can form the connection between boroxol rings in samples treated at higher temperatures. This configuration is also known to result from the transformation of tetrahedral boron ($BO_4$) to tetragonal boron ($BO_3$) by the formation of oxygen defects, specifically dangling bonds (or non-bridging oxygen (NBO)) and oxygen vacancies.[36,37]

As happens with a variety of oxides, heat treatment can promote the formation of oxygen vacancies that may, in turn, appear as neutral or charged defects.[38–41] These defects cause the formation of intragap levels in the oxide, which are responsible for absorption and emission in the near UV and visible.[42] Through EPR analysis, it has been observed that boric acid can generate additional defects in the form of radical species. These radicals can facilitate the absorption of radiation by promoting electron or trapped hole excitation.

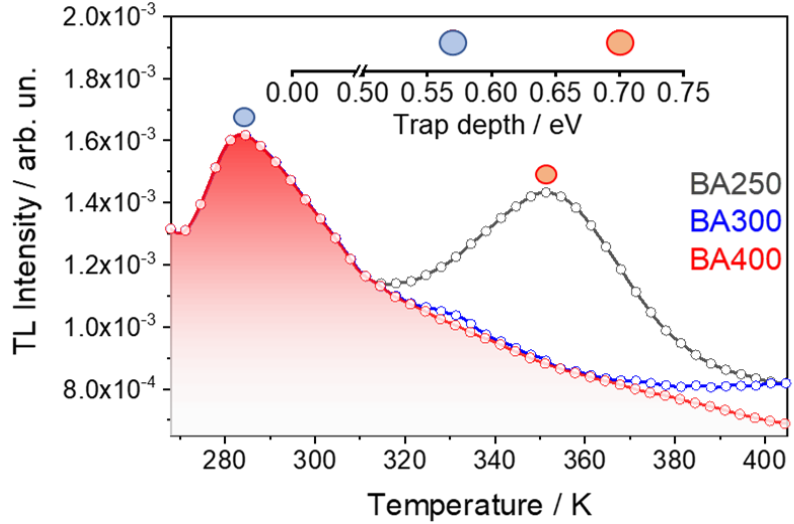

**Figure 6:** Thermoluminescence (TL) glow curves of BA250, BA300 and BA400 from 270 K to 400 K.

For this reason, we have explored the presence of trapped charge carrier by structural defects in BA250, BA300 and BA400 through thermoluminescence (TL) glow curves (**Figure 6**). BA250 exhibits a TL band with maximum at 349 K, which is not detectable in BA300 and BA400. This band is correlated to intragap levels that excite low-energy green emissions, which are removed after treatment at higher temperatures (300, 400 °C). The surviving blue emissions are characterized by lower traps observed at 283 K, which are common to all samples. Our findings suggest that the characteristics blue phosphorescence of treated boric acid is correlated to trapped charged carriers, whose shallow levels are estimated to be about 0.57 eV beneath the conduction band by the relation $E(eV) = T_m(K)/500$, where $T_m$ is the glow peak maximum.[43] To validate this approximation, we have also applied the generalized order of kinetic model (GOK)[43] described in the supplementary materials where all the parameters found for the glow curves are reported in **Table S2**.

Defect sites may delay the recombination at the optical center (color centers) due to trapping and de-trapping processes. In treated samples, thermal energy can empty the traps, favouring the non-radiative recombination at the color centers. The intrinsic defects that lead to this phenomenon may involve oxygen or boron vacancies as well as unsaturated bonds. Two traps are located at 0.57 and 0.70 eV from the bottom of the conduction band. Those levels can be filled up by light excitation and eventually emptied by thermal energy. The released electron can relax without emitting radiation and recombine at the optical centers. Low energy emission disappears after thermal treatment at 300 °C, demonstrating the correlation between color centers and traps. This effect can be attributed to trap healing or structural rearrangement. Accordingly, the phosphorescence in boric acid results from delayed recombinations driven by localized trap-like defects rather than spin prohibited transitions.[44]

Using the data obtained from the conducted experiments, we have constructed various representative structures depicting BOH coordination in different structural phases. Additionally, we have modelled probable local defects that could arise during thermal treatment, aiming to elucidate the nature of the defects referred to as BO$_x$ in EPR investigations. The resulting optimized geometries of stable molecules and clusters have been used to predict the absorbance spectra in defective structures to be correlated with the experimental excitation spectra.[45] Structural defects can be responsible for hole and electron trapping and determine intragap energy levels in oxides.[39,40,46–48]

Due to the insulating properties and weak hydrogen bonds of boric and metaboric acid, along with the glass produced through high-temperature treatment, employing small clusters and molecules becomes a viable approach for describing the diverse structural phases and exploring the factors influencing the formation of local structural defects. In **Figure 7** we display five exemplars of molecules with the corresponding calculated absorptions. **Figure 7a** illustrates the experimental excitation spectra as reference for absorbing channels in treated boric acid (BA250). The first edge vertical transition in the $H_3BO_3$ molecular unit (**Figure 7b**) is located at 173.79 nm (7.1342 eV), while the strongest absorptions are at 159.13 nm (7.7916 eV; $f$=0.0261) and 150.90 nm (8.2164 eV, $f$=0.0227) in good accordance with already reported simulations.[11] Simulated absorption of six $H_3BO_3$ molecules cluster shows a slight redshift of transitions up to 153.81 nm (8.06 eV) and the absorption edge located at 164.55 (7.53 eV) nm (**Figure S11**). However, there is no significant variation in the energy gap of the system, indicating that small clusters offer a reliable description for our specific structures. Similarly, calculated absorption in $H_3B_3O_6$ (**Figure 7c**) ring starts at 158.90 nm (7.8027 eV), compatible with $H_3BO_3$ structure. In general, although applied to simple models, the calculations support the absorption of $H_3BO_3$ at high energy in UV or far UV region and confirm previous reported DFT calculations on bulk boric acid and related polymorphs[30,31] indicating an indirect electronic bandgap higher than 6-7 eV. These values are a clear indication of the absence of near UV (NUV) transitions in absorption, which are supposed to trigger visible recombinations.

Defect-free bulk descriptions are inadequate to explain visible optical emissions and it is necessary to assume the formation of localized defects caused by atomic vacancies or charged molecular moieties. **Figures d)-e)-f)** show three examples of defective molecular clusters as a result of hypothesized intermediate structures after dimerization and trimerization of $H_3BO_3$ or $H_3B_3O_6$ rings. An extended survey of other 7 molecules and clusters is reported in **Table S3**.

According to the comparison between the theoretical absorption and the experimental spectra, bridging and non-bridging oxygen and oxygen vacancies are likely to promote UV and visible absorption in BOH structures. Figure 7d) illustrates the potential OOBO$^-$ defect as a result of a condensation of two $H_3BO_3$ units by dehydration process leading to the release of a water molecule. The resulting structure exhibits a non-bridging oxygen with an excess negative charge. The relative absorption spectrum shows a region that accurately reproduce the Urbach tail beyond 200 nm, along with an absorption band at 391.38 nm, compatible with the primary excitation channel.

Furthermore, the coordination of $BO_3$-$BO_4$ through bridging oxygen (Figure 7e) displays an absorption band at approximately 321 nm and multiple transitions in the blue region of the visible spectrum, with a peak around 440 nm, in excellent agreement with the excitation channels observed in PLE spectra. Additionally, the simulation of non-bridging oxygen in trigonal boron configuration that connects two $B_3O_3$ rings, as in the vitreous system, demonstrates a potential absorption at around 350 nm, consistent with the first excitation channel. In contrast, bond saturation inhibits the NUV and visible absorption as shown in cluster 6 of **Table S3**. Dangling oxygens or non-bridging oxygens can absorb in the visible range with a strong dependence on the charge of the cluster, i.e., the chemical potential of the material, as demonstrated by cluster 1 and 2 in **Table S3** that are able to alternatively reproduce the two excitation channels depending on their charge state.

Undercoordinated boron atoms in negatively charged cluster (BEC defect) is also responsible for the less energetic transitions that experimentally recombine by green emission (cluster 4, **Table S3**).

In our case, there is clear experimental evidence suggesting the impossibility of measuring transitions in defect-free boric acid and boron oxide structures. Recombinations in the visible spectrum result from defects in oxygen vacancies and non-bridging oxygen. They form over a limited temperature range during heat treatment, beyond which the oxidative process tends to quench luminescence. Simulations support this hypothesis, although it would be necessary to investigate the formation of said defects at grain boundaries and during melting along with the recrystallisation process, mechanisms that complicate the theoretical study of the process. The calculated experimental excitation and absorption spectra are in agreement and indicate the possibility of triggering both phosphorescent emissions with excitation energies close to the bandgap and with selective excitation in the near UV-visible regions.

Our experiments demonstrate an important aspect of the boric acid system. The $H_3BO_3$ structure is unlikely to exhibit any form of fluorescence or phosphorescence in its defect-free or impurity-free form. Moreover, the optical emission mechanism in the visible range in BA and its polymorphs is not interpretable with any form of through-space conjugation (TSC) between unpaired electrons of O atoms in confined space.[49] According to our interpretation, any emission from the BOH system must be attributed to the presence of defects that may occur during the synthesis process, aging, radiation exposure, and temperature fluctuations. In this context, although we do not detect phosphorescence in our samples before thermal processing, $H_3BO_3$ may manifest optical emissions assuming the existence of intragap recombination centers due to defectivity. This interpretation is consistent with our study on thermal treatment, revealing the occurrence of phosphorescence within a limited temperature range during the transformation process from crystal to amorphous state, involving dehydration, bond breaking, and condensation, leading to the formation of radicals, interstitial hydrogen, and oxygen vacancies. The presence of impurities, such as metallic ions or organic molecules, could also promote the stabilization of such defects, allowing the BOH matrix to release trapped charge carriers, inducing the phenomenon of phosphorescence in hybrid systems.

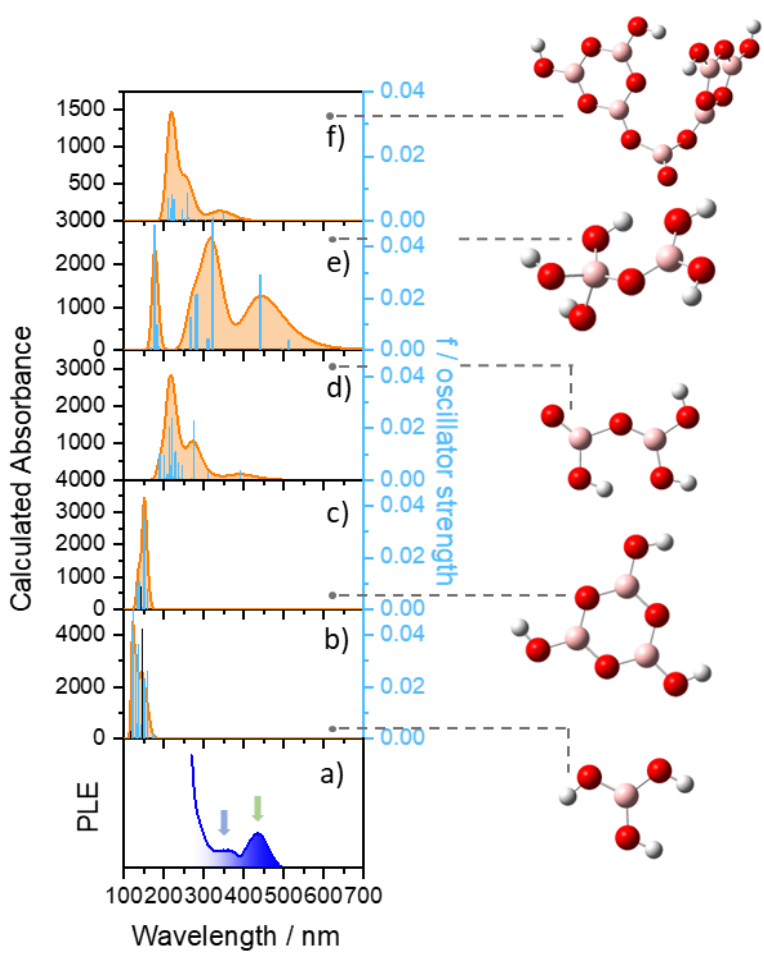

**Figure 7.** a) Experimental excitation bands from BA250 ($\lambda_{ex\_max,1}$ = 360 nm, $\lambda_{em\_max}$ = 480 nm; $\lambda_{ex\_max,2}$ = 435, $\lambda_{em\_max}$ = 528 nm and $\lambda_{ex,3}$ < 300 nm, $\lambda_{em\_max}$ = 480, 528 nm). b-f) calculated optical absorption and oscillator strength $f$ for the five different molecules and clusters: b) $H_3BO_3$ ($q$ = 0), c) $H_3B_3O_6$ ring ($q$ = 0), d) $H_3B_2O_5$ ($q$ = -1), e) $H_5B_2O_6$ ($q$ = 0), f) $H_4B_7O_{13}$ ($q$ = -1).

## 3. Conclusions

Heating boric acid to temperatures near or above its melting point results in structural defects that exhibit a phosphorescence when exposed to UV and visible light. Phosphorescence lasting hundreds of microseconds has been correlated to structural changes caused by a rise in temperature, specifically the transition to metaboric acid and boron oxide. In addition, the emission is characterized by two UV and visible excitation channels that relax in the blue and green regions, respectively.

The experimental results do not support the hypothesis that conventional spin forbidden recombinations in stoichiometric material or impurities are the source of phosphorescence in boric acid. Delayed recombination in presence of defective material is in better accordance with the experimental results.  The blue and green optical centres are compatible with the formation of structural defects of oxygen vacancies and unsaturated bonds. They are formed during the melting and recrystallisation process and reduce their occurrence by treatment at higher temperatures, where the oxide phase is predominant. The possibility of engineering defects as color centers in boric acid opens up new scenarios for the development of novel room-temperature phosphorescent materials.

## Author contributions

Supporting information

*for*

# Phosphorescence by trapping defects in boric acid induced by thermal processing

Luigi Stagi[1*], Luca Malfatti[1], Alessia Zollo[2], Stefano Livraghi[2], Davide Carboni[1], Daniele Chiriu[3], Riccardo Corpino[3], Pier Carlo Ricci[3], Antonio Cappai[3], Carlo Maria Carbonaro[3], Stefano Enzo[4], Abbas Khaleel[5], Abdulmuizz Adamson[5], Christel Gervais[6], Andrea Falqui[7], and Plinio Innocenzi[1,5*]

[1] Laboratory of Materials Science and Nanotechnology, CR-INSTM, Department of Biomedical Sciences, University of Sassari, Viale San Pietro 43/B, 07100 Sassari, Italy

[2] Department of Chemistry and NIS, University of Turin, Via P. Giuria 7, 10125 Turin, Italy

[3] Department of Physics, University of Cagliari. Sp 8, km 0.700, 09042 Monserrato, CA, Italy

[4] Department of Chemical, Physics, Mathematics and Natural Sciences, University of Sassari. Via Vienna 2. 07100 Sassari. Italy

[5] College of Science, Department of Chemistry. United Arab Emirates University. Al Ain. United Arab Emirates

[6] Sorbonne Université, CNRS, UMR 7574, Laboratoire de Chimie de la Matière Condensée de Paris, LCMCP, F-75005 Paris, France

[7] Department of Physics "Aldo Pontremoli", University of Milan, Via Celoria 16, 20133 Milan, Italy

Corresponding author email: lstagi@uniss.it ; plinio@uniss.it

**Experimental Section**

**Materials**

Boric acid (purity > 99.5 %) powders were purchased from Carlo Erba reagents s.r.l. (Milan). Commercial boric acid powders were treated at different temperatures for 2 h in air in an alumina crucible and cooled down under atmospheric conditions. Boron trichloride solution (1.0 M in methylene chloride) was purchased from Sigma-Aldrich.

**Synthesis of $H_3BO_3$ (BA-$H_3BO_3$*)**

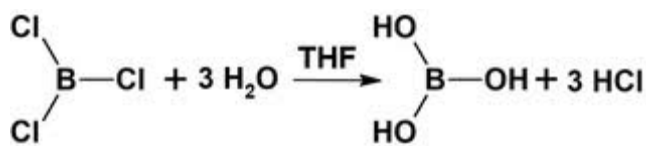

20 ml of THF were added to a 50 ml round-bottom flask equipped with a magnetic bar. 15 ml of boron trichloride solution ($BCl_3$, 1M in $CH_2Cl_2$, 15 mmol) were collected with a syringe, under nitrogen, from a sure-seal bottle and added to the THF that immediately turned from colourless to a pale blue solution. 3 ml of milli-Q water (3g, 167 mmol) were then carefully added to the THF solution under vigorous stirring (600 RPM) paying attention to the development of toxic and corrosive fumes (gaseous HCl) that started soon after the $BCl_3$ solution interacted with air moisture. After the water addition, a solid powder ($H_3BO_3$) started forming and precipitating. The resulting biphasic mixture (water-methylene chloride) was left to react under vigorous stirring for 2h to obtain boric acid, after which it was transferred to a 50ml plastic vial and centrifuged for 20 minutes at 7000 RPM. The surnatant was then removed and the solid was redispersed in 5ml milli-Q water and centrifuged again for 20 min at 7000 RPM. The surnatant was removed and the resulting off-white solid residue was recrystallised five times, from cold water in ice bath, until a snow-white crystalline solid was achieved.

**Characterization**

TGA-DSC analysis was performed using an SDTQ600 (TA instruments). All thermogravimetric analyses were done under a nitrogen ($N_2$) atmosphere.
A Shimadzu-6100 X-ray diffractometer using Cu-Kα radiation with a wavelength of 1.542 Å was used to obtain XRD patterns of the powder samples. The diffraction was measured within a range of 2θ angles from 10 – 50 °and recorded at a scan rate of 1 degree per minute.

Raman measurements were performed by a "Senterra" Raman microscope (Bruker) under a laser excitation of 785 nm (1 mW power). The spectra were collected with a resolution of ~3–5 cm$^{-1}$ and an integration time of 10 s. For highly fluorescent samples, Raman spectra were acquired by a fiber-coupled "BW-TEK: i-Raman® Plus" system operating under excitation at 1064 nm.

Fourier-transform infrared (FTIR) analysis was carried out with an infrared Vertex 70 interferometer (Bruker). FTIR spectra were recorded in the range 4000–400 cm$^{-1}$ with a 4 cm$^{-1}$ resolution.

*In-situ* Fourier-transform infrared spectra on sample powders in potassium bromide (KBr, IR 99%, Sigma) were collected in an electrical heating jacket in transmission geometry (Specac).

Micro X-ray Fluorescence (XRF) spectra were recorded by Micro-X ray Spectrofluorometer Bruker M4 Tornado in vacuum conditions (20 mbar) with X-ray tube operating at 50 keV and 600 μA.

Solid-state MAS NMR spectra were acquired on a Bruker AVANCE III 700 spectrometer with a 16.4 T superconducting solenoid operating at $v_0 = 224.68$ for $^{11}$B, using a commercial Bruker 3.2 mm MAS probe. Powder samples were transferred to $ZrO_2$ rotors and spun at a MAS rate of 20 kHz. $^{1}$H and $^{11}$B chemical shifts were referenced to TMS and $BF_3 \cdot OEt_2$, respectively.

Steady-state photoluminescence (PL) spectra were acquired in 3D fluorescence mapping mode [excitation (y)−emission (x)−intensity (z)] using a Horiba Jobin Yvon NanoLog spectrofluorometer equipped with a 450 W Xenon lamp as the excitation source. The maps were collected in the excitation/emission range of 250−650 nm with a 1 nm slit (increment = 5 nm) for excitation and emission.

Time-resolved phosphorescence measurements were performed by an optical parametric oscillator with a frequency doubler device, excited by the third harmonic (355 nm) of a pulsed Nd-YAG laser (Quanta-Ray Pro 730). The emission was dispersed by a spectrograph (ARC-SpectraPro 300i) with a spectral bandpass <2.5 nm and detected by a gateable intensified CCD (Princeton Inst.).

Continuous wave EPR spectra were acquired using a Bruker EMX spectrometer operating at X-band (9.5 GHz), equipped with a cylindrical cavity using a 100 kHz field modulation. All the spectra were recorded in vacuum conditions at 77 K in an EPR cell that can be connected to a conventional high-vacuum apparatus (residual pressure < 10$^{-4}$ mbar), with 1 mW of incident microwave power and a modulation amplitude of 0.2 mT. *In-situ* irradiation experiments with UV-Vis light were carried out using a 500 W Xe-Hg lamp (covering the range UV-Vis-IR, from ~ 250 nm to IR wavelength (> 2400 nm)).

Thermally stimulated luminescence (TSL) measurements were carried out, after sample irradiation for five minutes at 250 nm with a deuterium lamp (16 W). The emission was monitored in a temperature range between 260–420 K. The experiment was executed with a home-made set-up consisting of a power heater controlled by an Agilent 6030A System Power Supply coupled with an automatic PID loop multichannel HP 3852A Data Acquisition/Control Unit which recorded the thermocouple signal of the sample temperature. The measure was set up with a rate of 0.16 K s$^{-1}$ and the TSL signal was collected with a photomultiplier EMI-9816 QB and registered in a second channel of the same HP unit. A dedicated LabView2019 software was realized to control the instruments automatically. The glow curves were analyzed by the generalized order kinetic model (GOK), where the thermal response can be described by the superposition of n components:

$$I(T) = \sum_{j=1}^{n} A_j S_j exp\left(-\frac{E_j}{kT}\right)\left[1 + \frac{(b_j-1)S_j}{\beta}\int_{T_1}^{T_2} exp\left(-\frac{E_j}{kT}\right)dT\right]^{-\frac{b_j}{b_j-1}}$$

where A is a proportionality constant, which depends on the traps concentration, E is the thermal trap depth (the thermal energy needed to release the trapped charge carrier that can be different from the optical energy needed to release the trapped carrier), k is Boltzmann's constant, b and S are phenomenological parameters related to the trapping kinetics and frequency factor respectively, and β is the sample heating rate. Second order kinetics (or bimolecular) typically arises when there is a high probability of re-trapping the released electrons with respect to the recombination at the luminescent center while the frequency factor S (or ''attempt-to-escape'' frequency) has the physical meaning of the number of times per second that a bound electron interacts with the lattice phonons Table S2 reports all the obtained parameters.

Quantum Chemistry calculations were carried out using Gaussian 16 code.[2] . In order to verify the possible role of defects in UV/vis absorption spectra of boric acid structures different guess structures were generated and their geometries were optimized at DFT/ωB97XD/6-311+G(d,p) level of theory to capture the influence of dispersion-related contributions (included in the formulation of the functional) and the long-range nature of molecular orbitals (via the diffusion functions explicitly added in the basis set).[3] UV/vis absorption spectra were eventually calculated on the relaxed structures in the framework of Time-Dependent DFT (TD-DFT) using the B3LYP/6-311+G(d,p) combination of basis set and functional.[4,5] Analysis of frequencies confirms that optimized structures are at a minimum of potential surface, and no imaginary frequencies were obtained. GaussView 6 was used to interpret the computed data.[6]

**Figures and Tables**

**Figure S1.** a) Rietveld refinement for the estimation of the occurrence of $H_3BO_3$ and $HBO_2$ in samples BA, BA100, BA130 and BA200 t. b) Relative composition (in %) of boric acid, metaboric acid ($HBO_2$) and boron oxide ($B_2O_3$) after each temperature treatment. Estimation of degree of crystallinity has been determined by calculating $A_c / (A_a + A_c)$ where $A_a$ is the area of under broad bands and $A_c$ is the area of crystalline sharp peaks in XRD patterns. c) Average dimensions (in nm) of the boric acid and metaboric acid crystallites.

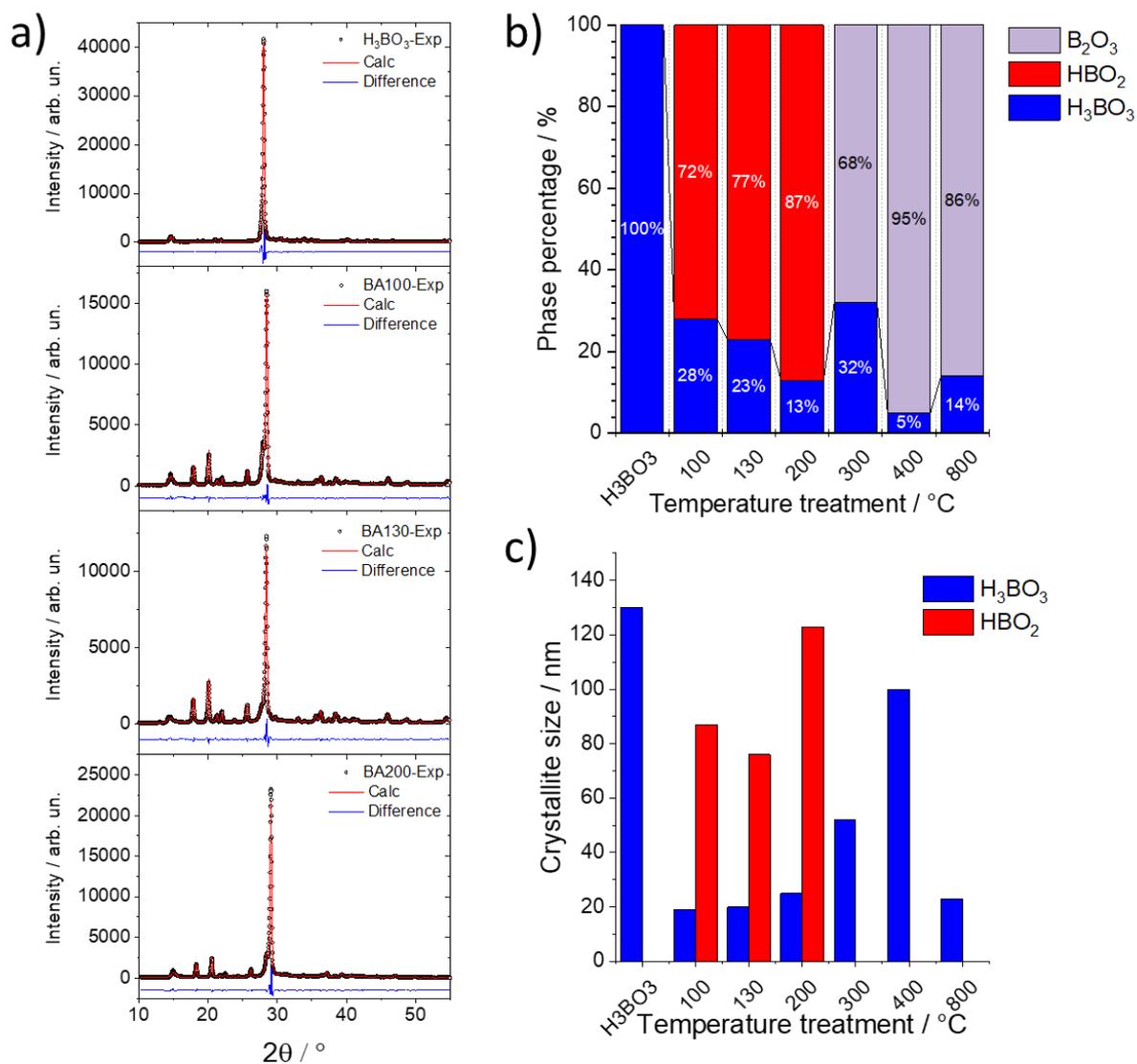

**Figure S2:** a) Raman spectra of boric acid (BA, $H_3BO_3$) and b) boric acid treated at 800 °C (BA800). c) Attribution of Raman active frequencies in samples BA and BA800. d) Pictures of vibrational displacements direction.

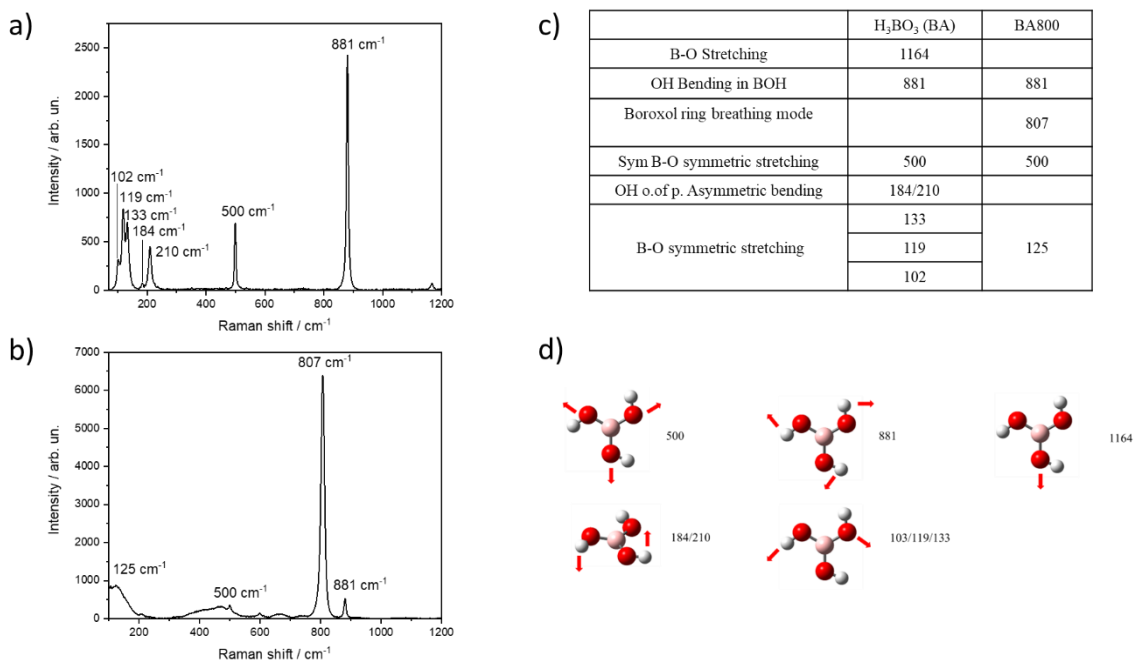

**Figure S3**: FTIR absorption spectra of boric acid treated at 100, 200, 300, 400, 500, 600 and 800 °C.

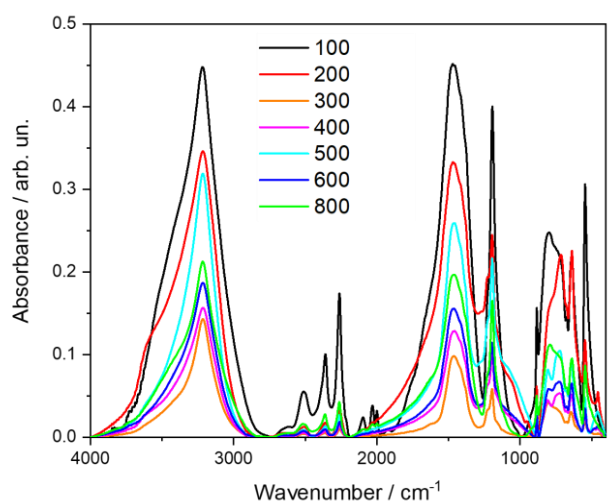

**Figure S4**: a) ¹¹B ad b) ¹H experimental MAS NMR spectra of boric acid and BA samples.

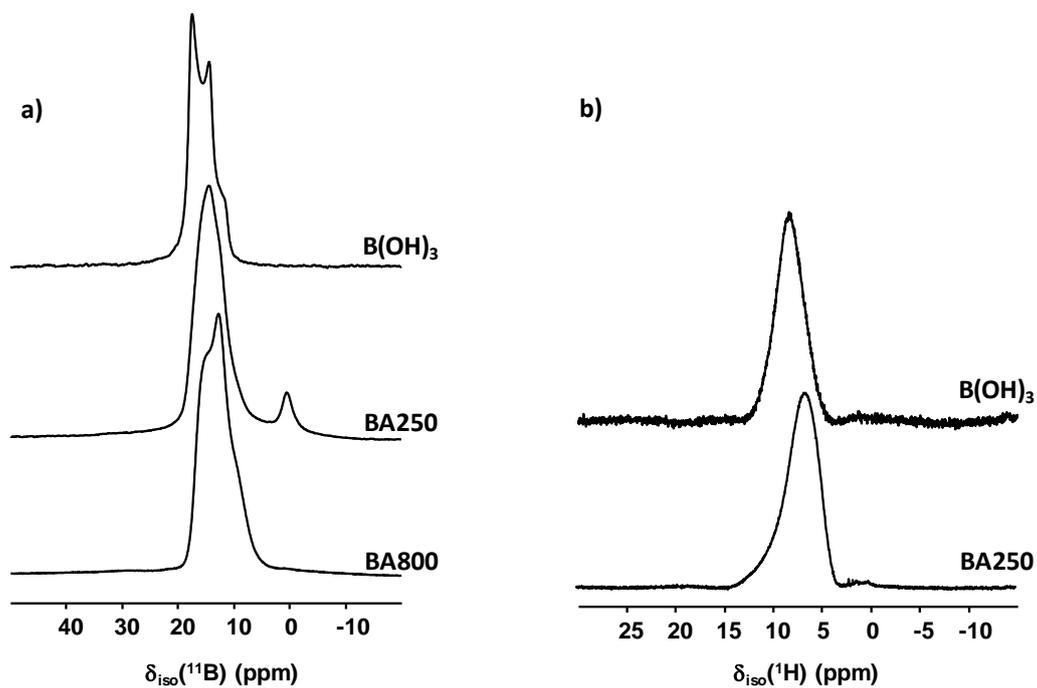

**Figure S5**: Excitation spectra (PLE) of BA250 in the range between 200 and 500 nm. Inset: detail of PLE between 300 and 500 nm.

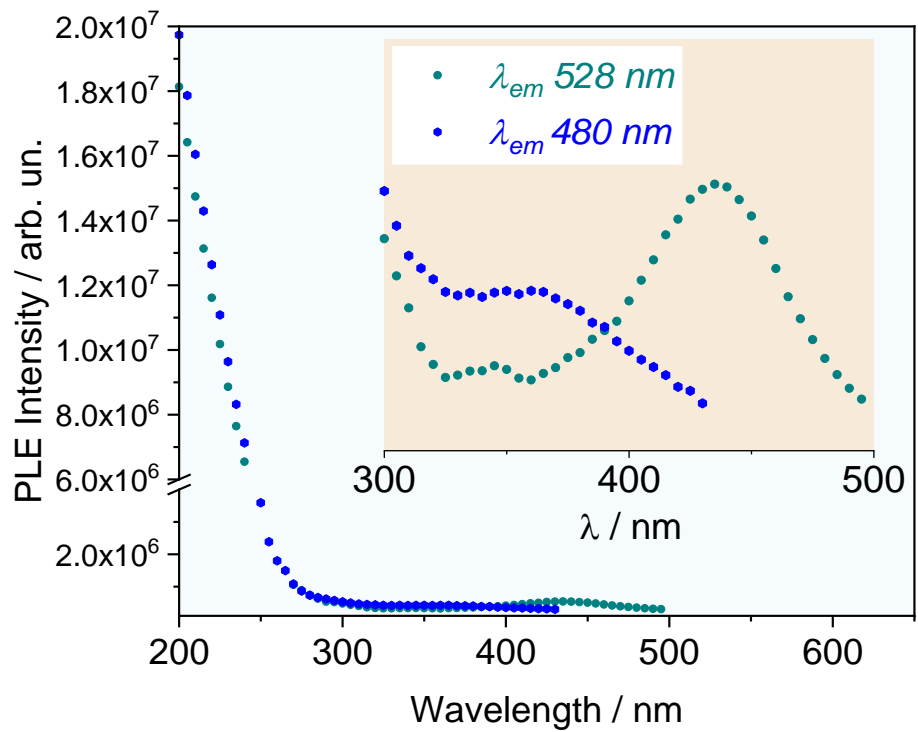

**Table S1**: Fitting parameters (two-exponential decay) of time resolved luminescence profile.

| | $I = I_1\exp(-t/\tau_1) + I_2\exp(-t/\tau_2) + I_0$<br>$I_1, I_2$: Intensity (arb.un.)<br>$\tau$: lifetime (ms)<br>$\chi^2$ : chi-sq. | |
|---|---|---|
| | EM: 410 nm (BA200); 480 nm (BA250, BA300, BA400) | EM 520 nm (BA200), 528 nm (BA250, BA300, BA400) |
| $I_1$ | 0.72 | 0.68 |
| $\tau_1$ (ms) | 0.632 | 0.426 |
| $I_2$ | 0.28 | 0.32 |
| $\tau_2$ (ms) | 1.656 | 1.278 |
| $\chi^2$ | 0.996 | 0.998 |

**Figure S6**: a) image of boric acid treated at 100 °C for 1 h on silica substrate. b) image of boric acid treated at 200 °C for 1 h on silicon substrate and point dependent Raman spectra.

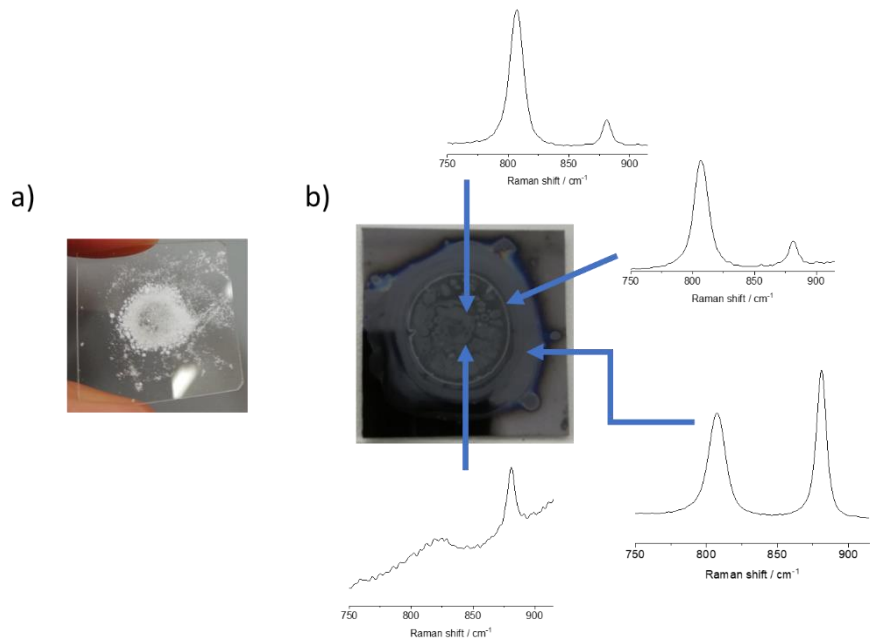

**Figure S7**: a) XRD diffraction pattern of BA-H$_3$BO$_3$* obtained from BCl$_3$ after synthesis process. b) excitation and emission spectra of the synthesized sample after thermal treatment at 250 °C. c) Time-resolved phosphorescent emissions monitored at 480 and 525 nm.

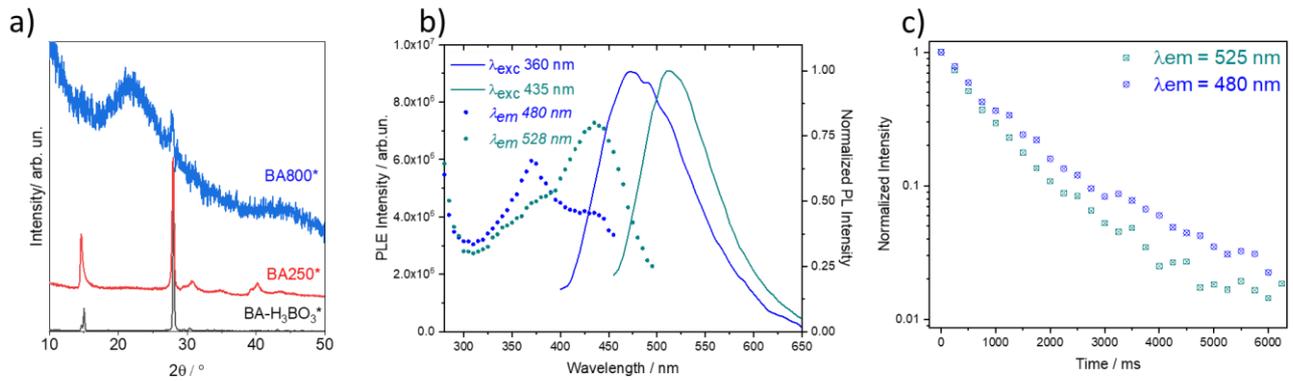

**Figure S8**: XRF spectra of four representative samples in the ranges 1-5 eV and 1-8 eV (extended). Al, S, and Ca Kα lines serve as indications of the expected energy of selected elements.

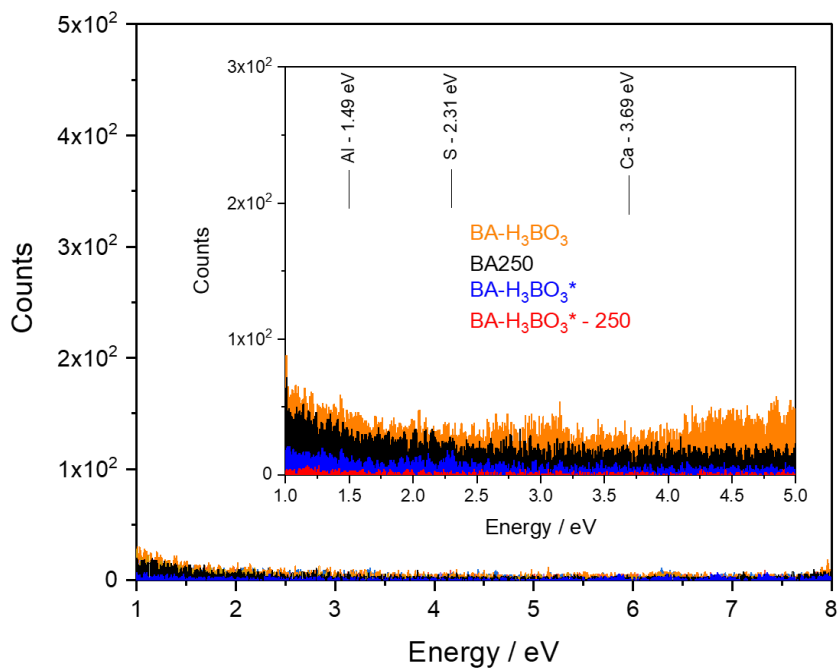

**Figure S9:** Magnification of the EPR signal centred at g = 2003 (Centre 1) reported in *Figure 5* of the main text. The stick diagram highlights the hyperfine splitting attributable to the interaction of the unpaired electron with the boron nucleus. The observed splitting is reasonable with the presence of a nucleus with nuclear spin I = 3/2 as in the case of the most abundant isotope of boron ($^{11}$B). In this case a splitting in four components is expected in according to the formula **n° lines = 2nI+1** (where n = number of equivalent nuclei with nuclear spin I)

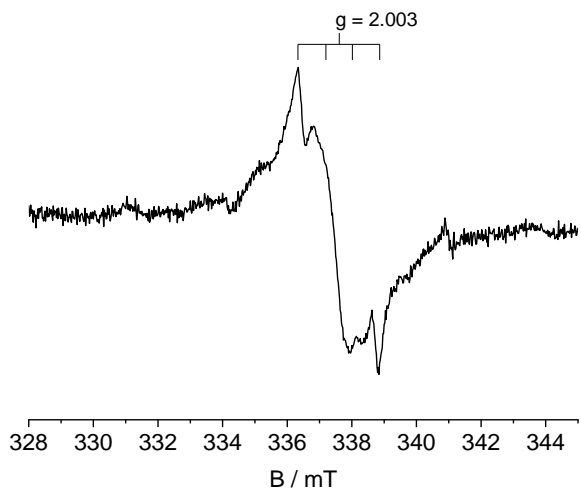

**Figure S10:** Comparison of the EPR spectra obtained under UV irradiation for the sample BA250 and the pristine $H_3BO_3$.

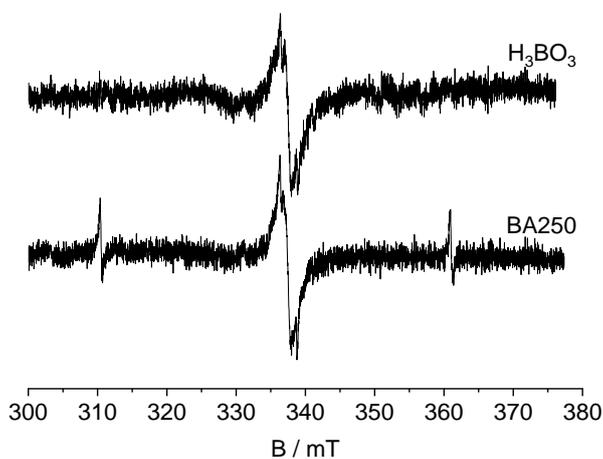

**Table S2: Fitting parameters of Thermoluminescence glow curves.**

| Trap 1 | | | | Trap 1 | | | | Heating rate |
|---|---|---|---|---|---|---|---|---|
| *A1* | *b1* | *S1(s-1)* | *E1 (eV)* | *A2* | *b2* | *S2(s-1)* | *E2 (eV)* | *B (K/s)* |
| 0.062 | 1.003 | 1.64E+08 | 0.564 | 0.103 | 1.05 | 8.74E+07 | 0.69 | 0.2 |

**Figure S11**: Left ) highest occupied molecular orbital (HOMO) and lowest unoccupied molecular orbital representation. Right) calculated UV-Vis absorption of 6(H$_3$BO$_3$) cluster and orbitals involved in the strongest vertical transition.

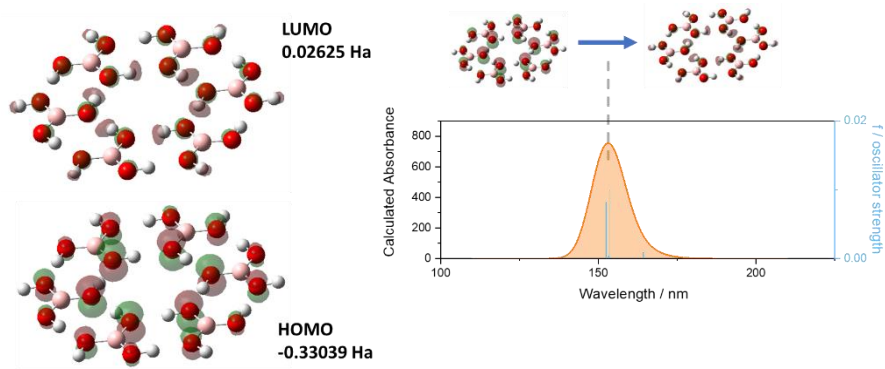

**Table S3**: Structures and corresponding simulated absorption of different B-O-H molecules and clusters.

| # | Structure | Calculated UV-Vis absorption | Charge *Spin* |
|---|---|---|---|
| 1 | 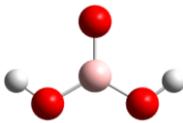 | 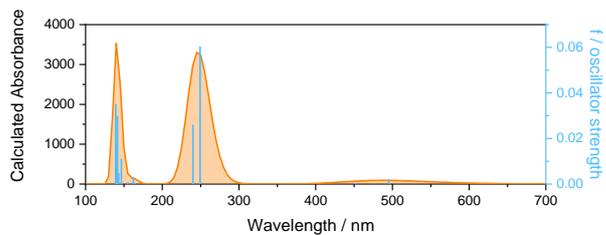 | Q = 0; *doublet* |
| 2 | 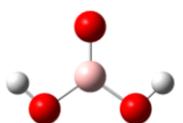 | 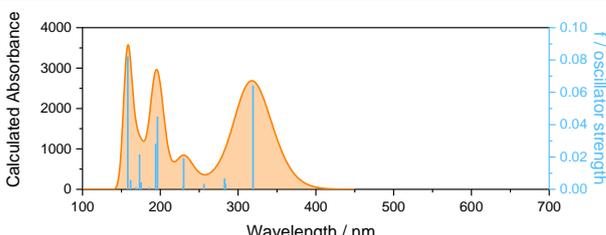 | Q = +1e; *singlet* |
| 3 | 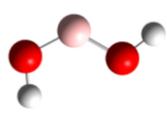 | 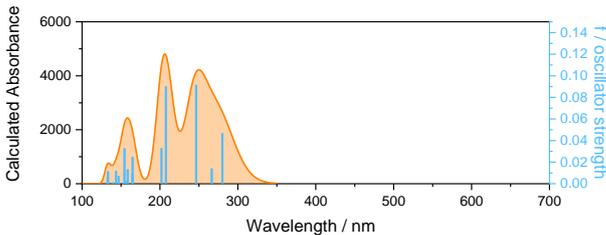 | Q = 0; *doublet* |
| 4 | 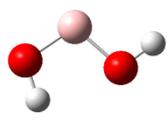 | 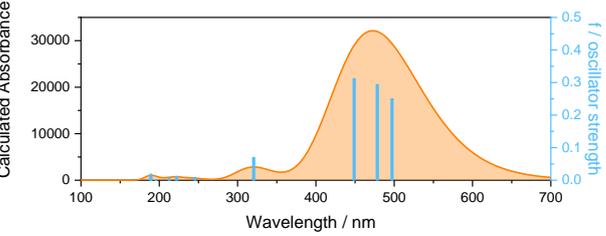 | Q = +1e; *singlet* |
| 5 | 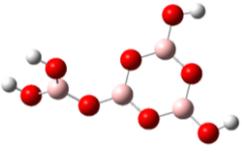 | 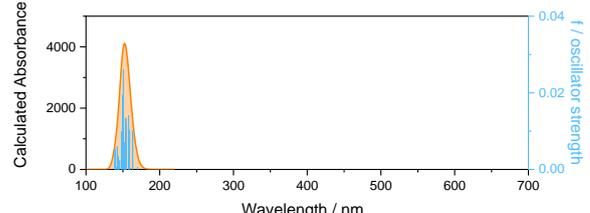 | Q = 0; *singlet* |
| 6 | 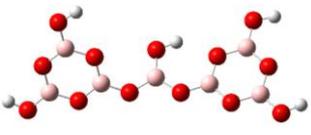 | 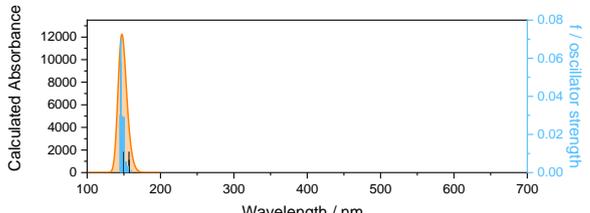 | Q = 0; *singlet* |
| 7 | 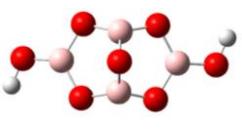 | 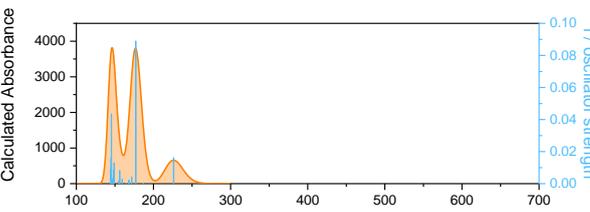 | Q = 0; *singlet* |